\documentclass[aps,twocolumn,preprintnumbers,amsmath,amssymb,footinbib,prl]{revtex4}
\usepackage{epsfig}
\usepackage{graphicx}
\usepackage{bm}
\newcommand{\be}{\begin{equation}}
\newcommand{\ee}{\end{equation}}
\newcommand{\bea}{\begin{eqnarray}}
\newcommand{\eea}{\end{eqnarray}}

\newcommand{\citenamefont}{}
\newcommand{\bibfnamefont}{}
\newcommand{\bibnamefont}{}

\begin{document}
\title{Nuclear Matter and Neutron stars in a Parity Doublet Model}
\author{V.~Dexheimer$^1$, S.~Schramm$^{1,2}$, D.~Zschiesche$^3$}
\affiliation{$^{1}$ Franfurt Institute for Advanced Studies,
J.W.\ Goethe Universit\"at,\\
Ruth-Moufang-Str. 1, 60438 Frankfurt, Germany}
\affiliation{$^{2}$ Center for Scientific Computing,
J.W.\ Goethe Universit\"at,\\
Max-von-Laue Str.\ 1, 60438 Frankfurt, Germany}
\affiliation{$^{3}$ Institut f\"ur Theoretische Physik,
J.W.\ Goethe Universit\"at,\\
Max-von-Laue Str.\ 1, 60438 Frankfurt, Germany}

\begin{abstract}

We investigate the properties of isospin-symmetric nuclear matter and neutron stars
in a chiral model approach adopting the SU(2) parity doublet formulation. This ansatz explicitly incorporates chiral symmetry restoration
with the limit of degenerate masses of the nucleons and their parity
partners. Instead of searching for an optimized parameter set we explore the general parameter dependence of nuclear
matter and star properties in the model. We are able to get a good description of ground state nuclear matter as well
as large values of mass for neutron stars in agreement with observation.

\end{abstract}

\maketitle

\noindent
\section{Introduction}

The use of effective hadronic models based on chiral symmetry has been quite successful in describing
nuclear matter, finite nuclei and neutron stars (see e.g. \cite{model}).
While most of the approaches are based on variations of the linear sigma model or its non-linear realization,
work on an alternative ansatz starting from the parity-doublet model (\cite{Jido:1998av,Nemoto:1998um,Kim:1998up})
has been quite limited.
In this approach, the scalar sigma meson serves to split the nucleon and its parity partner, while in the chirally restored
phase both baryonic states become degenerate but not massless. This situation is quite analogous to
sigma models with hidden gauge symmetry \cite{Bando} where the masses of vector and axial vector mesons are split in a similar way to
attain degenerate (but non-zero) masses in the chirally restored phase similar to the pion and sigma mesons.
It has only quite recently been shown \cite{ZParity} that with this approach one can obtain saturated nuclear matter with relatively high nuclear
compressibilities with a simple potential for the scalar field and a cross-over transition to chiral restoration at large densities.
In our investigation we calculate symmetric nuclear matter and neutron star properties within this framework.
In order to get a better understanding of the parameters involved and not just to fix an arbitrary fine-tuned
parameter set we vary the parameters over a broad range of values and calculate nuclear matter properties and star masses.

The parameter scan shows that
for quite a wide range of values a satisfactory reproduction of
nuclear matter binding energy and saturation density is possible.
Further demanding that the nuclear matter compressibility lies within
a range of reasonable values (assumed to be between 100 and 400 $\rm{MeV}$ in the following), restricts the parameter space
significantly but still allows for a range of values that yield a quantitatively good description of nuclear matter
ground state properties in contrast to the results in \cite{ZParity}.
In order to study isospin asymmetric nuclear matter the rho meson is included in the Lagrangian. Its coupling to the nucleons is
fixed by the phenomenological value for the asymmetry energy of about 32 MeV.
By further requiring that the maximum mass of the neutron star, determined by solving the Tolman-Oppenheimer-Volkov (TOV)
equations, should have a mass around $1.9$ $\rm{M_solar}$, we find additional
restrictions for the allowed model parameters.

The article is organized as follows. The parity doublet model is briefly introduced in the following section.
We then apply the equations to symmetric nuclear matter and to
isospin-asymmetric matter as it is present in neutron stars. The
properties of the model are first tested to fit saturation
quantities and then to reproduce observational properties such as the
maximum mass of the neutron star. Finally, we discuss the results of this investigation
and add some remarks on further work to be done.

\section{Model Description}

In the parity doublet model the nucleons and their parity partners
belong to the same multiplet and so are true chiral partners \cite{Jido:1998av,Nemoto:1998um,Kim:1998up}.
Under $SU_L(2) \times SU(2)_R$ transformations $L$ and $R$,
the two nucleon
fields $\psi_1$ and $\psi_2$ transform as:
\begin{eqnarray}
\psi_{1R} \longrightarrow R \psi_{1R} \  & , \hspace{1cm} &   \psi_{1L}
\longrightarrow L \psi_{1L} \ , \label{mirdef1} \\
\psi_{2R} \longrightarrow L \psi_{2R} \  & , \hspace{1cm} &   \psi_{2L}
\longrightarrow R \psi_{2L} \ . \label{mirdef2}
\end{eqnarray}
This allows for a chirally invariant mass, $m_0$:
\begin{eqnarray}
&&m_{0}( \bar{\psi}_2 \gamma_{5} \psi_1 - \bar{\psi}_1
      \gamma_{5} \psi_2 ) = \nonumber \\
&& m_0 (\bar{\psi}_{2L} \psi_{1R} -
        \bar{\psi}_{2R} \psi_{1L} - \bar{\psi}_{1L} \psi_{2R} +
        \bar{\psi}_{1R} \psi_{2L}) \ . \label{chinvmass}
\end{eqnarray}
The full chiral Lagrangian reads:
\begin{eqnarray}
{\cal L} &=& \bar{\psi}_1 i {\partial\!\!\!/} \psi_1
+ \bar{\psi}_2 i {\partial\!\!\!/} \psi_2 \nonumber\\
&+& m_0 \left(\bar{\psi}_2 \gamma_5 \psi_1 - \bar{\psi}_1 \gamma_5
  \psi_2\right)\nonumber\\
&+& a \bar{\psi}_1 \left(\sigma + i \gamma_5 \boldsymbol{\tau}
  \cdot\boldsymbol{\pi}\right) \psi_1
+ b \bar{\psi}_2 \left(\sigma - i \gamma_5 \boldsymbol{\tau}
  \cdot\boldsymbol{\pi}\right) \psi_2\nonumber\\
&-& g_{\omega} \bar{\psi}_1 \gamma_{\mu} \omega^{\mu} \psi_1
- g_{\omega} \bar{\psi}_2 \gamma_{\mu} \omega^{\mu} \psi_2 \nonumber \\
&-& g_{\rho} \bar{\psi}_1 \gamma_{\mu}\boldsymbol{\tau}\cdot\boldsymbol{\rho}^{\mu} \psi_1
- g_{\rho} \bar{\psi}_2 \gamma_{\mu}\boldsymbol{\tau}\cdot\boldsymbol{\rho}^{\mu} \psi_2 \nonumber \\
&+& {\cal L}_M \ ,
\label{lagrangian}
\end{eqnarray}
where $a$, $b$, $g_{\omega}$ and $g_{\rho}$ are the coupling constants of the mesons
fields  ($\sigma$, $\pi$, $\omega$ and $\rho$) to the baryons $\psi_1$
and $\psi_2$.  Note that we assume the same vector coupling strength for
both parity partners, as it would result from minimal coupling of
the vector mesons. The
mesonic Lagrangian ${\cal L}_M$ contains the kinetic terms
of the different meson
species, and potentials for the scalar and vector fields.
The potential for the spin zero fields is the same as in the ordinary
SU(2) linear sigma model.
Kinetic and potential terms are added for an isoscalar vector
meson, $\omega$,
as in the $\sigma$-$\omega$ model of nuclear matter \cite{Walecka:1974qa} as well as for the isovector vector meson, $\rho$:
\begin{eqnarray}
{\cal L}_M&=&\frac{1}{2} \partial_{\mu} \sigma^{\mu} \partial^{\mu}
\sigma_{\mu}
+ \frac{1}{2} \partial_{\mu} \vec{\pi}^{\mu} \partial^{\mu}
\vec{\pi}_{\mu}
- \frac{1}{4} \omega_{\mu \nu} \omega^{\mu \nu} \nonumber \\
&-& \frac{1}{4} \boldsymbol{\rho}_{\mu \nu} \boldsymbol{\rho}^{\mu \nu}
+ \frac 12 m_\omega^2 \omega_{\mu} \omega^{\mu} + \frac{1}{2} m_\rho^2 \boldsymbol{\rho}_{\mu} \boldsymbol{\rho}^{\mu}
 \nonumber \\
&+&g_4^4 [(\omega_{\mu}\omega^{\mu})^2+(\boldsymbol{\rho}_{\mu} \boldsymbol{\rho}^{\mu})^2+6 (\omega_{\mu}\omega^{\mu}\boldsymbol{\rho}_{\mu} \boldsymbol{\rho}^{\mu})]
\nonumber\\
&+& \frac 12 \bar{\mu}\,^2 (\sigma^2+\vec{\pi}^2) - \frac \lambda 4
(\sigma^2+\vec{\pi}^2)^2  \nonumber \\
&+& \epsilon\sigma \ ,
\end{eqnarray}
where $\omega_{\mu \nu}=\partial_{\mu}
\omega_{\nu}-\partial_{\nu}\omega_{\mu}$ and $\boldsymbol{\rho}_{\mu \nu}=\partial_{\mu}
\boldsymbol{\rho}_{\nu}-\partial_{\nu}\boldsymbol{\rho}_{\mu}$ represent
the field strength tensors of the vector fields.
As usual, the parameters $\lambda$, $\bar{\mu}$ and $\epsilon$
can be related to the sigma and pion masses, and the pion decay
constant, in vacuum:
\begin{eqnarray}
\lambda&=&\frac{m_{\sigma}^2-m_{\pi}^2}{2 \, \sigma_0^2} \nonumber \ ,\\
\bar{\mu}\,^2&=&\frac{m_{\sigma}^2-m_{\pi}^3}{2} \nonumber \ ,\\
\epsilon&=&m_{\pi}^2 f_{\pi} \ ,
\end{eqnarray}
with $m_{\pi}=138$ $\rm{MeV}$, $f_{\pi}=93$ $\rm{MeV}$ and the  vacuum expectation value of
the sigma field $\sigma_0=f_{\pi}$.
Since the mass of the $\sigma$ meson
in the vacuum can not be fixed precisely by experiment, we will treat
it as a free parameter.
The vacuum mass of the $\omega$ field is $m_{\omega}=783$ $\rm{MeV}$ and the vacuum mass of the $\rho$ field is $m_{\rho}=761$ $\rm{MeV}$.

To investigate the properties of dense nuclear matter and the chiral
phase transition at zero temperature, we adopt the mean-field approximation
\cite{Serot:1984ey}. The fluctuations around constant vacuum expectation
values of the mesonic field operators are neglected, while the nucleons
are treated as quantum-mechanical one-particle operators.  Only the
time-like component of the isoscalar vector meson $\bar{\omega} \equiv
\omega_0 $ and the time-like third component of the isovector vector meson $\bar{\boldsymbol{\rho}} \equiv
\ \rho_{03} $ survive, assuming  homogeneous and isotropic infinite
nuclear matter.  Additionally, parity conservation demands $\bar{\boldsymbol{\pi}}
=0$. The mass eigenstates for the parity doubled nucleons, the $N_+$ and $N_-$ are determined by diagonalizing  the mass matrix, Eq.~(\ref{chinvmass}), for $\psi_1$ and $\psi_2$. Writing the coupling constants a and b as a function of $M_{N_+}(939 \rm{MeV})$, $M_{N_-}$ and $\sigma_0$,
the effective masses of the baryons are given by:

\begin{eqnarray}
{M_N^*}_\pm&=&\sqrt{\left[\frac{(M_{N_+}+M_{N_-})^2}{4}-m_0^2\right]\frac{\sigma^2}{\sigma_0^2}+m_0^2}\nonumber\\
&\pm&\frac{M_{N_+}-M_{N_-}}{2}\frac{\sigma}{\sigma_0}\ ,
\end{eqnarray}

If chiral symmetry is completely restored, i.e. $\sigma=0$,
the two nucleonic parity states become degenerate in mass with
${M_N^*}_+={M_N^*}_-=m_0$. Thus, for large values of $m_0$,
the sigma field mainly serves to split the nucleon and its parity partner
nucleon masses and contributes less to the (dynamical) mass of the nucleon ~\cite{Jido:1998av}.
Independent of the value for $m_0$, in the vacuum, i.e. for $\sigma=\sigma_0$, the baryon effective masses reproduce their measured values.

The thermodynamic potential per unit volume is:
\begin{eqnarray}
\frac{\Omega}{V}=-{\cal L}_M+\sum_i \frac{\gamma_i}{(2 \pi)^3}
\int_{0}^{k_{F_i}} \,
d^3k \, (E_i^*(k)-\mu_i^*) \ ,
\end{eqnarray}
where $i \in \{N_+,N_-\}$ denotes the nucleon type,
$\gamma_i$ is the fermionic degeneracy,
$E_{i}^* (k) = \sqrt{k^2+{M^*_i}^2}$ the energy,
and $\mu_i^*=\mu_i-g_{\omega} \omega_0-g_{\rho} \rho_{03} \tau_3/2=\sqrt{k_F^2+{M^*_i}^2}$
the corresponding effective chemical potential.
The single
particle energy of each parity partner $i$ is given by
$E_i(k)=E_i^*(k)+g_{\omega} \omega_0+g_{\rho} \rho_{03} \tau_3/2$.

The mean meson fields $\bar{\sigma}$, $\bar{\omega}$  and $\bar{\rho}$ are determined by extremizing the
thermodynamic potential $\Omega/V$:
\begin{eqnarray}
\left . \frac{\partial(\Omega/V)}{\partial \sigma} \right |_{\bar{\sigma},\bar{\omega},\bar{\rho}}
&=&-\bar{\mu}\,^2 \bar{\sigma} +
\lambda \bar{\sigma}^3 - \epsilon \ ,\nonumber \\
&+& \sum_i {\rho_s}_i(\bar{\sigma},\bar{\omega},\bar{\rho})
\left . \frac{\partial M^*_i}{\partial  \sigma} \right |_{\bar{\sigma}} =0 \ ,\nonumber \\
\left . \frac{\partial(\Omega/V)}{\partial \omega_0} \right |_{\bar{\sigma},\bar{\omega},\bar{\rho}} &=&-m_\omega^2
\bar{\omega} - g_4^4 (4\bar{\omega}^3+12\bar{\omega}\bar{\rho}^2)\ ,\nonumber \\
&+&g_{\omega} \sum_i \rho_i (\bar{\sigma},\bar{\omega},\bar{\rho})=0 \ , \nonumber \\
\left . \frac{\partial(\Omega/V)}{\partial \rho_0^3} \right |_{\bar{\sigma},\bar{\omega},\bar{\rho}} &=&-m_\rho^2
\bar{\rho} - g_4^4 (4\bar{\rho}^3+12\bar{\omega}^2\bar{\rho})\ , \nonumber \\
&-&g_{\rho} \sum_i \left(\rho_i^n
(\bar{\sigma},\bar{\omega},\bar{\rho})-\rho_i^p(\bar{\sigma},\bar{\omega},\bar{\rho})\right)=0 \ . \nonumber \\
&& \ \
\label{eqsmotion}
\end{eqnarray}
The scalar density ${\rho_s}_i$ and the baryon density $\rho_i$ for each
chiral partner are given by the usual expressions:
\begin{eqnarray}
  {\rho_s}_i&=&\gamma_i \int_0^{k_{F_i}} \frac{d^3k}{(2 \pi)^3}
  \,\frac{M^*_i}{E^*_i}\nonumber \\
  &=&\frac{\gamma_i M^*_i}{4 \pi^2}
  \left[k_{F_i} E_{F_i}^*-{M^*_i}^2 {\rm ln} \left(
      \frac{k_{F_i}+E_{F_i}^*}{M^*_i} \right) \right] \ , \nonumber \\
  \rho_i&=&\gamma_i \int_0^{k_{F_i}} \frac{d^3k}{(2
    \pi)^3}=\frac{\gamma_i k^3_{F_i}}{6 \pi^2} \ .
\label{densities}
\end{eqnarray}

The basic nuclear matter saturation properties
we impose can be formulated in the following way. At the point of vanishing pressure we have
\begin{eqnarray}
\label{nucmatprop}
E/A-M_N &=& -16.0 \pm 0.5\rm{MeV} \ , \nonumber \\
\rho_0 &=& 0.15 \pm 0.015 \mbox{ fm}^{-3}  \ .
\end{eqnarray}
Note that in contrast to former work in \cite{ZParity}, here we allow for
a reasonable range of values for the different observables.

\section{Results}

\subsection{Nuclear Matter}

\begin{figure}[h]
\centering
\includegraphics[width=9.0cm]{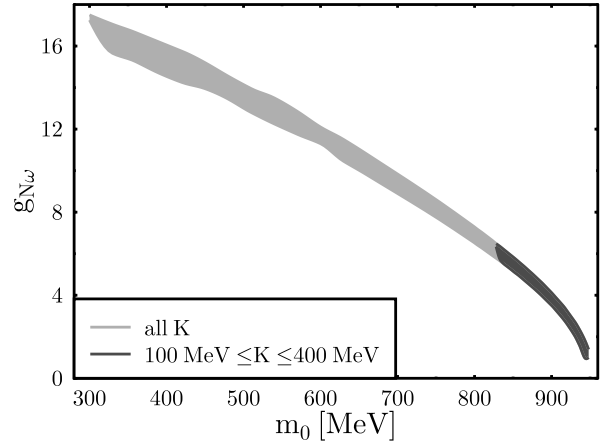}
\vspace{-0.6cm}
\caption{\label{gnovsm0}
Value of the nucleon vector coupling constant $g_{N\omega}$
vs mass parameter $m_0$.}
\end{figure}

\begin{figure}[h]
\centering
\includegraphics[width=9.0cm]{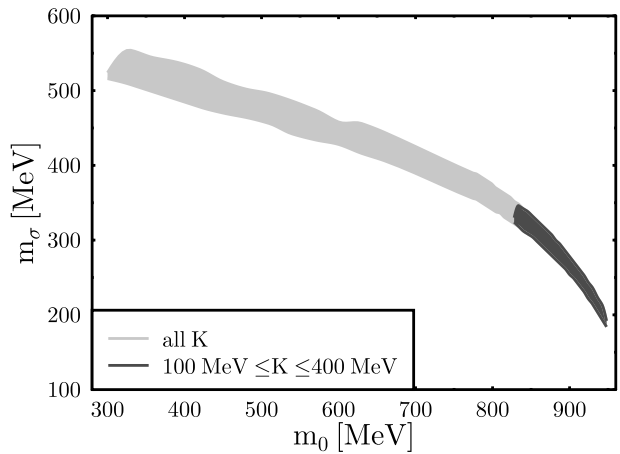}
\vspace{-0.6cm}
\caption{\label{msvsm0}
Mass of sigma meson in vacuum vs mass parameter $m_0$.}
\end{figure}

First we apply the model to symmetric nuclear matter and determine the ranges of the
parameters that reproduce the saturation properties
for binding
energy per nucleon and baryon density as given by
eq. \ref{nucmatprop}. In principle the free parameters of
the model are the coupling
constant of the
vector meson to the nucleons $g_{N\omega}$, the mass of the chiral partners $M_{N_-}$ (as the identification
with a specific experimentally known resonance is not completely unambiguous),
the bare mass $m_0$, the coupling constant of the fourth order self interactions of
the vectorial
fields $g_4$ and the mass of the scalar meson $m_\sigma$ as already
described in \cite{ZParity}.
However, to keep things simple, we will for most of this work
restrict our analysis to parameter sets with $g_4=0$ and  $M_{N_-}=1535$ $\rm{MeV}$.
Where necessary or interesting we will discuss the influence of finite
values of $g_4$ or smaller values of $M_{N_-}$, as was also done
in \cite{ZParity}.
The other parameters are varied over all possible values.
In Fig. (\ref{gnovsm0}) and Fig. (\ref{msvsm0}) we show the values of the
corresponding parameters vs mass parameter $m_0$, for which the
imposed nuclear matter saturation properties are fulfilled.
In both figures, the light grey bands represent the allowed regions
without any further restriction, while the dark grey bands depict the
parameter sets, which in addition yield a reasonable nuclear matter
compressibility  $K=9 \frac{\partial P}{\partial \rho_i}=9 \rho_i \frac{\partial^2 \epsilon}{\partial \rho_i^2}=
9\rho_i \frac{\partial \mu_B}{\partial \rho_i}$ at saturation between 100 and 400 MeV.
As one can see, successful fits without restricting the
compressibility are possible for $300 \lesssim m_0 \lesssim 950$~\rm{MeV}.
This region is larger than that found in \cite{ZParity}
because no fine-tuning to fixed value for the nuclear matter saturation properties has been
done here.
The corresponding nucleon-vector coupling constant $g_{N\omega}$ as well as the
value of the sigma mass in vacuum $m_\sigma$ decrease with increasing
values of $m_0$, as has also been observed in \cite{ZParity}.

Demanding the nuclear matter
compressibility to be in ''reasonable ranges'' restricts the possible parameter values. For our choice of
$100 \mbox{ MeV} < K < 400 \mbox{ MeV}$, we see that only the fits
with quite large $m_0$ values ($> 800$~\rm{MeV}) survive.
\begin{figure}[h]
\centering
\includegraphics[width=9.0cm]{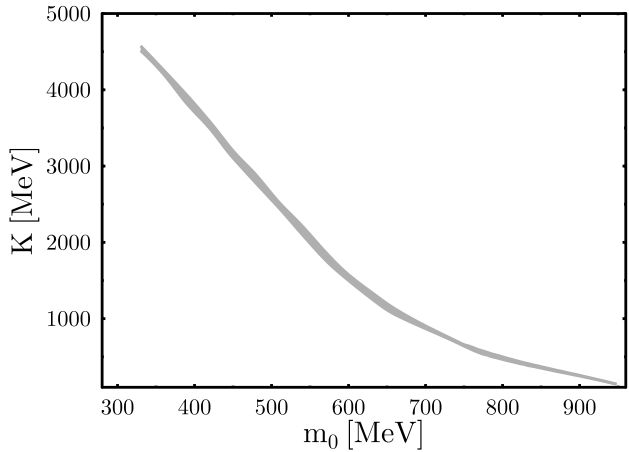}
\vspace{-0.6cm}
\caption{\label{Kvsm0}
Nuclear compressibility at saturation vs mass parameter $m_0$.}
\end{figure}

This can be seen in more detail from Fig. (\ref{Kvsm0}). For smaller
values of $m_0$ the nuclear matter compressibility increases very
fast beyond any acceptable value.

\subsection{Asymmetric Matter and Neutron Stars}

\begin{figure}[h]
\centering
\vspace{-1.0cm}
\includegraphics[width=9.0cm]{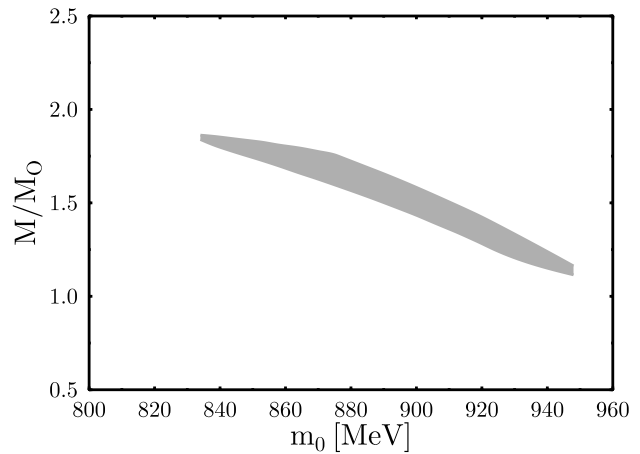}
\caption{\label{nstarmass}
Mass of neutron stars vs mass parameter $m_0$.}
\end{figure}

In order to study neutron star properties we calculate the equation of state (EOS)
of charge neutral matter including leptons.
Due to the different densities of neutrons and protons,
the expectation value of the vector-isovector meson $\rho$  does not
vanish. The corresponding coupling constant is determined
such that an asymmetry energy at saturation of $E_{sym} = 32.5$
$\rm{MeV}$ is obtained.
Neutron star masses and radii are determined by solving the Tolman-Oppenheimer-Volkov
equations
\cite{mass,mass2}. The maximum star mass for a given equation of state is shown as a function of $m_0$ in Fig. (\ref{nstarmass})
for the cases that have a compressibility at saturation $K$ between
$100$ and $400$ $\rm{ MeV}$. As can be seen, the maximum
mass of the star decreases with increasing $m_0$. This results from the fact that
for larger bare mass term $m_0$, medium effects are reduced which leads to a smaller
compressibility as can be seen in fig. \ref{Kvsm0} and reduced
repulsion at high densities and consequently lower star masses.
Thus, demanding the compressibility to be in a reasonable range,
especially not to be too big and in addition, to
demand to have a maximum star mass above a phenomenologically acceptable value of at least 1.4 to
1.6 solar masses, restricts the value of $m_0$ to be somewhere
between 900 and 870 MeV. Thus the initially broad band
of possible parameter sets is strongly narrowed by these additional
observables. Note, that these values of $m_0$ are quite large, which
means that the mass generation is very much different from a standard
linear $\sigma$ model, where the mass is more or less exclusively
generated by the dynamical symmetry breaking, while now most of the
mass is generated by the parity-mixing mass term.
The results for neutron stars show that keeping a high mass for the
chiral partners and a small coupling constant of the fourth order self interactions
of the vectorial fields produce more massive stars, being in agreement
with phenomenology.

\begin{figure}[h]
\centering
\vspace{-1.0cm}
\includegraphics[width=9.0cm]{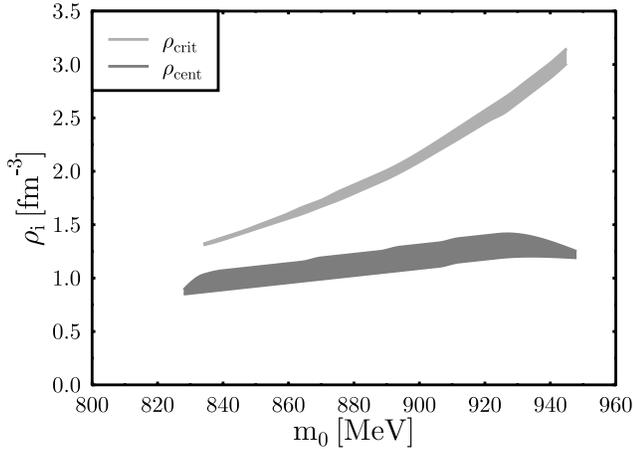}
\caption{\label{nstarden} Central densities of neutron stars vs mass
parameter $m_0$ and critical density for chiral restoration or density
for appearance of N-,
respectively.}
\end{figure}

\begin{figure}[h]
\centering
\vspace{-1.0cm}
\includegraphics[width=8.cm]{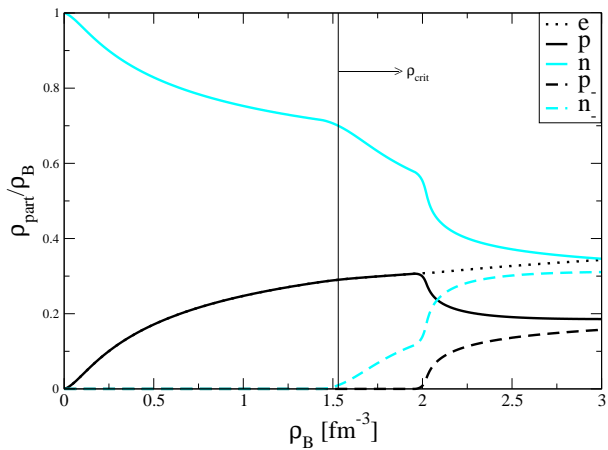}
\caption{\label{pop1}
Population of the star vs baryon density for $m_0=843$ $\rm{MeV}$ and $M_{N_-}=1535$ $\rm{MeV}$. The central density of the respective star would be $\rho_{cent}=0.96 fm^{-3}$.}
\end{figure}

\begin{figure}[h]
\centering
\vspace{-1.0cm}
\includegraphics[width=8.0cm]{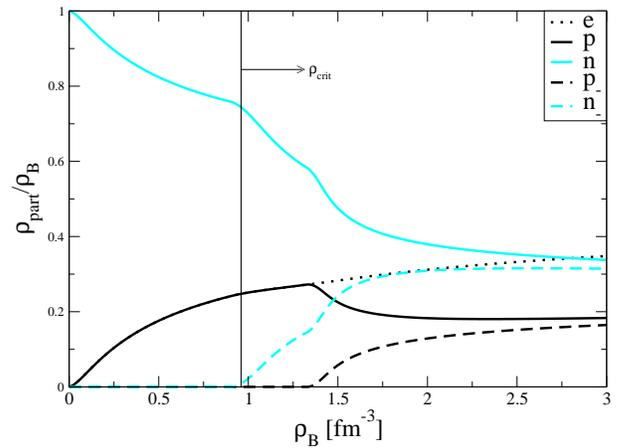}
\caption{\label{pop2}
Population of the star vs baryon density for $m_0=855$ $\rm{MeV}$ and $M_{N_-} =1379$ $\rm{MeV}$. The central density of the respective star would be $\rho_{cent}=1.04 fm^{-3}$.}
\end{figure}

\begin{figure}[h]
\centering
\vspace{1.0cm}
\includegraphics[width=8.0cm]{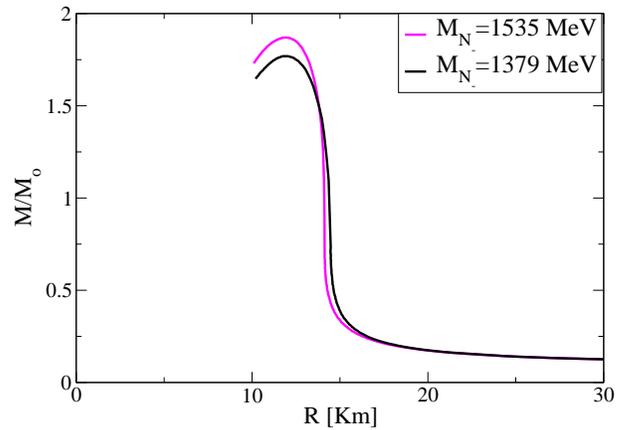}
\caption{\label{MassCrust}Mass radius diagram for different chiral partners masses.}
\end{figure}

In figure \ref{nstarden}, the central densities of the neutron stars
and the threshold densities above which the $N_-$ states are populated, respectively, are shown.
The maximum mass stars have a typical central energy density of around $1fm^{-3}$ which for
the cases at hand with $g_4=0$ and $M_{N_-} = 1535$~MeV
is always smaller than the critical density to create chiral partners
and
in this case no $N_-$ are present in the star as can be seen in the
particle densities shown in Fig. \ref{pop1}.
However as has already been observed in \cite{ZParity}, decreasing the
mass of the $N_-$, reduces the threshold density and thus can allow for the population of $N_-$ states in
neutron stars.
This can for example be seen in Fig. \ref{pop2}, for
$M_{N_-}=1379$ $\rm{MeV}$, where the chiral partners
become populated already slightly below $\rho_B = 1 \mbox{ fm}^{-3}$.
The mass-radius relation for the two cases is
shown in Fig.
(\ref{MassCrust}).
Besides the EOS for the dense part of the star, the EOS for an outer crust, an
inner crust and an atmosphere
have been added \cite{?}.

$\newline$

As can be seen, the maximum mass of the star is higher for
$M_{N_-}=1535$ $\rm{MeV}$ ($M_{max}=1.87$ $\rm{M_o}$)
compared with the case for $M_{N_-}=1379$ $\rm{MeV}$ ($M_{max}=1.77$
$\rm{M_o}$), as expected.
However, both masses are around $1.9$ $\rm{M_o}$,
value that is in agreement with the massive stars observed lately with
the heaviest one so far having a mass of $M=2.1^{+0.4}_{-0.5}$ $\rm{M_o}$ \cite{21}.

$\newline$

$\newline$

$\newline$

$\newline$

$\newline$

\section{Conclusions}

We applied the parity doublet model to nuclear and neutron star matter.
We were able to achieve a quantitatively good description of saturated nuclear matter
as well as to obtain maximum masses of neutron stars that are in agreement with recent observations.
In our approach we explored a wide range of model parameters that
reproduce saturation properties of nuclear matter for a window of values for the
density ($\rho_0=0.15fm^{-3}\pm10\%$) and binding energy per nucleon ($E/A-M_N=-16\pm0.5\rm{MeV}$).
With that we could study the dependence of results on the free parameters of the model:
the coupling constant of the vector meson to
the nucleons $g_{N\omega}$, the bare mass $m_0$ and the mass of the scalar meson $m_\sigma$. Initially
the mass of the chiral partners is assumed to be $M_{N_-}=1535\rm{MeV}$, corresponding to one of the likely candidates listed in the particle
data book. No vector meson self-interactions have been taken into account as they generally lead to very soft equations of state
and small masses of neutron stars that are in disagreement with observation. Besides that, if we
take into account phenomenological values of compressibility (allowing for a somewhat wider range $100\rm{MeV} < K < 400\rm{MeV}$)
we find that the mixing parameter $m_0$  has to be $m_0 > 800 \rm{MeV}$
so that the two different couplings of the nucleon to the scalar field
cancel each other to a large degree.

In a second step the asymmetry between protons and neutrons has been included demanding a charge neutral state (including the electron)
and the $\rho$ meson coupling constant is determined to reproduce the phenomenological value for the asymmetry energy at saturation.
If we consider a high mass for the nucleon chiral partners $M_{N_-}$, the critical density above which those particles appear in the star is also high,
and in the case of $M_{N_-}=1535\rm{MeV}$ they do not appear in the star. The value of the maximum mass is $M_{max}=1.87\rm{M_0}$.
Demanding that the $M_{max}$ should be at least 1.4 solar masses restricts the value of the explicit mass parameter $m_0$ to be less than 900 MeV.
Decreasing the mass of the chiral partners by about $150\rm{MeV}$ is enough to decrease the threshold density of the
chiral partners for them to be present in the star, which reduces the maximum mass of the star moderately,
in the case of $M_{N_-}=1379\rm{MeV}$ to $M_{max}=1.77$ $\rm{M_o}$. An interesting extension of this work would be the generalization
to a Flavor-SU(3) approach. Work in this direction is in progress.

\section{Acknowledgement}
We thank G. Pagliara and J. Schaffner-Bielich for valuable discussions. The numerical work has been done at the
computing facilities of the CSC Frankfurt.

\end{document}